\documentclass[aip,apl,preprint,floatfix,amsfonts,amsmath,amssymb]{revtex4-2}

\usepackage{graphicx,dcolumn,bm}

\begin{document}

\title{Origin of negative anomalous Nernst thermopower in Mn-Ga ordered alloys}

\author{Weinan Zhou}
\email{ZHOU.Weinan@nims.go.jp}
\affiliation{Research Center for Magnetic and Spintronic Materials, National Institute for Materials Science (NIMS), Tsukuba 305-0047, Japan}

\author{Keisuke Masuda}
\affiliation{Research Center for Magnetic and Spintronic Materials, National Institute for Materials Science (NIMS), Tsukuba 305-0047, Japan}

\author{Yuya Sakuraba}
\email{SAKURABA.Yuya@nims.go.jp}
\affiliation{Research Center for Magnetic and Spintronic Materials, National Institute for Materials Science (NIMS), Tsukuba 305-0047, Japan}
\affiliation{PRESTO, Japan Science and Technology Agency, Saitama 332-0012, Japan}

\begin{abstract}
The negative sign of the anomalous Nernst thermopower ($S_\text{ANE}$) observed in Mn-Ga ordered alloys is an attractive property for thermoelectric applications exploiting the anomalous Nernst effect (ANE); however, its origin has not been clarified.
In this study, to gain insight into the negative $S_\text{ANE}$, we prepared epitaxial thin films of Mn$_{x}$Ga$_{100-x}$ with $x$ ranging from 56.2 to 71.7, and systematically investigated the structural, magnetic, and transport properties including the anomalous Hall effect (AHE) and the ANE.
The measured $S_\text{ANE}$ is negative for all samples and shows close to one order of magnitude difference among different compositions.
Together with the measured transport properties, we were able to separate the two different contributions of the ANE, i.e., one originating from the transverse thermoelectric coefficient ($\alpha_{xy}$), and the other one originating from the AHE acting on the longitudinal carrier flow induced by the Seebeck effect.
Both contributions are found to be negative for all samples, while the experimentally obtained negative $\alpha_{xy}$ exhibits a monotonic increase towards zero with increasing $x$, which is consistent with the tendency indicated by first-principles calculations.
Our results show that the large difference in the negative $S_\text{ANE}$ is mostly attributed to $\alpha_{xy}$, and thus shed light on further enhancement of the ANE in Mn-based ordered alloys.
\end{abstract}

\maketitle

The anomalous Nernst effect (ANE) refers to the emergence of a transverse electric field in a magnetic material perpendicular to both the temperature gradient ($\nabla{T}$) and magnetization.
In contrast to the Seebeck effect (SE) having a parallel relationship between the electric field and $\nabla{T}$, the ANE allows different design principles for thermoelectric power generation and heat flux sensing applications, which could realize a simpler and more flexible device with better integratability and scalability.\cite{TTG1,TTG2,TTG3,TTG4,TTG5,TTG6,TTG7,TTG8,TTG9,TTG10,TTG11}
In a similar manner to a $\Pi$-shaped junction used in a thermoelectric generator consisting of p-type and n-type semiconductor pellets, a junction made by connecting two magnetic wires having positive and negative anomalous Nernst thermopower ($S_\text{ANE}$) could be the basic unit for the ANE-based devices.
The generated transverse electric fields from the two magnetic wires are towards opposite directions when their magnetization is aligned to the same direction and under the same $\nabla{T}$, which means the output is added up if the wires are connected at one end to form a junction.
Such a junction could then be duplicated and connected in series to form a simple two-dimensional meander structure for thermoelectric power generation or heat flux sensing.\cite{TTG2,TTG4,TTG10}
$S_{\text{ANE}}$ can be expressed as
\begin{equation}
S_{\text{ANE}} = \rho_{xx}\alpha_{xy} - \rho_\text{AHE}\alpha_{xx}, \label{eq1}
\end{equation}
where $\rho_{xx}$, $\rho_\text{AHE}$, $\alpha_{xx}$, and $\alpha_{xy}$ are the longitudinal resistivity, anomalous Hall resistivity, longitudinal thermoelectric coefficient, and transverse thermoelectric coefficient, respectively.\cite{eqSANE1,eqSANE2}
The first term on the right hand side of Eq.~(\ref{eq1}), $\rho_{xx}\alpha_{xy}$ (defined as $S_{\text{I}}$), originates from $\alpha_{xy}$ directly converting $\nabla{T}$ into a transverse electrical current as $j_{y} = \alpha_{xy}\nabla{T}$.
The second term on the right hand side of Eq.~(\ref{eq1}), $- \rho_\text{AHE}\alpha_{xx}$ (defined as $S_{\text{II}}$), originates from the anomalous Hall effect (AHE) acting on the longitudinal carrier flow induced by the SE, and can be rewritten as $- S_\text{SE} \times \rho_\text{AHE} / \rho_{xx} = - S_\text{SE} \times \text{tan}(\theta_\text{AHE})$.
In recent years, it is shown that $\alpha_{xy}$ is closely linked to the Berry curvature of the electronic bands of materials, and the estimation of $\alpha_{xy}$ through the ANE measurement could be used to probe the topological band structures, leading to an increasing interests in studying the ANE from the viewpoint of fundamental physics in addition to thermoelectric applications.\cite{axy1,axy2,axy3,axy4,axy5,axy6,axy7,axy8,axy9,axy10,axy11,axy12,axy13}
Large values of $S_\text{ANE}$ have been reported, i.e., in $L$2$_{1}$-Co$_{2}$MnGa Heusler alloy, which was attributed to its substantial $\alpha_{xy}$.\cite{CMG1,CMG2,CMG3,CMG4}
The ANE of Mn-Ga ordered alloys, on the other hand, generates a transverse electric field in the opposite direction (a negative $S_\text{ANE}$).\cite{TTG2,SANEMnGa2}
Such a property could play an complementary role in the meander structure for the thermoelectric applications exploiting the ANE.
Comparing to connecting magnetic wires with non-magnetic electrodes,\cite{TTG8} the use of magnetic materials having negative $S_\text{ANE}$ could lead to larger output and higher efficiency.

\begin{figure}
\includegraphics[width=8.6cm]{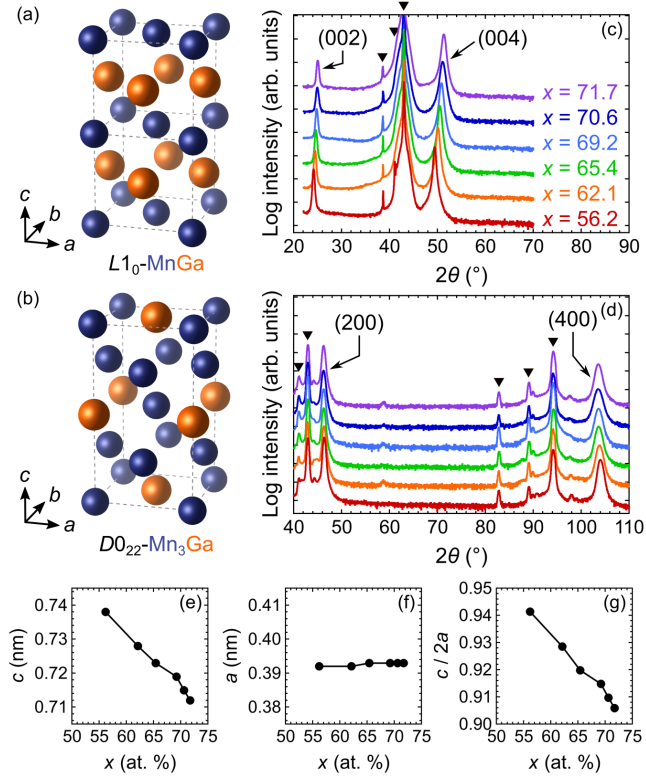}
\caption{\label{Fig1} Schematic illustrations of the crystal structure for (a) $L1_{0}$-MnGa and (b) $D0_{22}$-Mn$_{3}$Ga. The unit cell for $L1_{0}$-MnGa in (a) is doubled along the $c$ axis for comparison. (c) Out-of-plane and (d) in-plane XRD patterns for the Mn$_{x}$Ga$_{100-x}$ films. The symbol $\blacktriangledown$ marks the diffraction peaks from the single crystal MgO (100) substrate due to Cu $K_{\alpha}$, $K_{\beta}$, and W $L_{\alpha}$ radiation. (e) Out-of-plane lattice constant $c$ and (f) in-plane lattice constant $a$ estimated using the results in (c) and (d), respectively, as a function of the atomic percent of Mn $x$. (g) Tetragonal distortion ratio $c/2a$ as a function of $x$.}
\end{figure}

Mn-Ga ordered alloys have been popular as candidate materials for spintronics applications due to their unique properties.\cite{MnGa1,MnGa2,MnGa3,MnGa4,MnGa5}
For Mn$_{x}$Ga$_{100-x}$, the $L1_{0}$ ordering [Fig.~\ref{Fig1}(a)] is thermodynamically stable when $x \sim$ 50$-$65;
with increasing atomic percent (at.~\%) of Mn to $x \sim$ 65$-$75, the $D0_{22}$ ordering appears [Fig.~\ref{Fig1}(b)], where two kinds of Mn sites couple antiferromagnetically.
The Mn-Ga alloys are known to have large uniaxial magnetic anisotropy ($K_\text{u}$), and relatively small saturation magnetization ($M_\text{s}$) compared to common magnetic materials, such as Fe, Co, or their alloys.
These properties would be beneficial when exploiting the ANE for thermoelectric applications,\cite{TTG8} if the Mn-Ga alloys also have a large $|S_\text{ANE}|$.
However, a clear understanding of the reason behind the negative $S_\text{ANE}$ is currently lacking.
A systemic investigation on how the composition would affect $S_\text{ANE}$ is also desirable, since such knowledge may provide hints for further enhancement of the ANE in Mn-Ga ordered alloys.

In this study, through both experiments and first-principles calculations, we investigated the ANE in Mn-Ga ordered alloys.
We prepared epitaxial Mn-Ga thin films with different compositions and characterized their ANE.
The obtained values of $S_\text{ANE}$ are negative for all samples, however, exhibit close to one order of magnitude large difference among different compositions.
Combining with the experimentally measured transport properties, we were able to disentangle the $S_{\text{I}}$ and $S_{\text{II}}$ terms of the ANE, where both were found to be negative.
In addition, the large difference in the negative $S_\text{ANE}$ is mostly originated from $S_{\text{I}}$.
Experimentally obtained $\alpha_{xy}$ is negative for all samples and shows a monotonic increase towards zero with increasing $x$.
This behavior is consistent with the tendency indicated by $\alpha_{xy}$ of $L1_{0}$-MnGa and $D0_{22}$-Mn$_3$Ga obtained from first-principles calculations.

The Mn-Ga thin films were deposited on MgO (100) single crystal substrates by magnetron sputtering.
Most of the Mn-Ga films were formed by co-deposition from a Mn target and a Mn$_{50}$Ga$_{50}$ target, with the compositions controlled by varying the power to the cathodes; only the sample with the least Mn was deposited from a single Mn$_{54}$Ga$_{46}$ target.
For all the samples, the same thermal treatment process was used.
Prior to the deposition, the surface of the substrate was cleaned by heating the MgO substrate to 600 $^\circ$C for 30 minutes.
Then, the substrate temperature was kept at 400 $^\circ$C for the deposition of 50-nm-thick Mn-Ga film, followed by post-annealing at 500 $^\circ$C for 1 hour.
When the sample was cooled to room temperature, 3-nm-thick Al capping layer was deposited to prevent oxidation.
In total, 6 samples with different compositions of Mn$_{x}$Ga$_{100-x}$ were prepared, with $x =$ 56.2, 62.1, 65.4, 69.2, 70.6, and 71.7, which were determined by wavelength dispersive X-ray fluorescence analysis.
The structure of the Mn-Ga films was studied using out-of-plane and in-plane X-ray diffraction (XRD) with Cu $K_{\alpha}$ radiation.
The magnetic properties were measured with a vibrating sample magnetometer.
In order to characterize the transport properties and the thermopower, the Mn-Ga films were patterned into a 2-mm-wide and 7-mm-long Hall bar structure using photolithography and Ar ion milling, followed by the formation of Au electrodes using a lift-off process.
The longitudinal resistivity and anomalous Hall resistivity of the Mn-Ga films were measured using a physical property measurement system (PPMS).
The measurements for $S_\text{SE}$ and $S_\text{ANE}$ were carried out using a home-made holder embedded in a multi-function probe, which was used with the PPMS.
A Peltier module inside the holder was used to apply temperature gradient to the Mn-Ga films.
Details of the measurement for obtaining $S_\text{SE}$ and $S_\text{ANE}$ are described in a previous paper.\cite{STTG1}
All the measurements were performed at room temperature.

Using first-principles calculations, $\sigma_{xy}$ and $\alpha_{xy}$ of the stoichiometric $L1_0$-MnGa and $D0_{22}$-Mn$_3$Ga alloys were theoretically studied.
First, the electronic structures of these alloys are calculated using the density-functional theory (DFT) with the full-potential linearized augmented plane-wave method including the spin-orbit interaction, which is implemented in the WIEN2k program.\cite{2020Blaha-JCP}
We adopted the generalized gradient approximation\cite{1996Perdew-PRL} for the exchange-correlation energy in the DFT calculation.
The lattice constants $a=$ 3.92 (3.93) \AA~and $c=$ 3.69 (7.12) \AA~were used for the calculations in $L1_0$-MnGa ($D0_{22}$-Mn$_3$Ga), which were estimated based on the results of the XRD measurement.
Using the obtained electronic structures, we calculated $\sigma_{xy}$ given by\cite{2004Yao-PRL}
\begin{eqnarray}
\sigma_{xy}(\epsilon)&=&-\frac{e^2}{\hbar} \int \frac{d^3k}{(2\pi)^3}\,\, \Omega^{z}({\bf k}),\label{eq2}\\
\Omega^{z}({\bf k})&=&- {\left(\frac{\hbar}{m}\right)}^2\, \sum_{n} f(E_{n,{\bf k}},\epsilon) \sum_{n' \neq n} \frac{2\,{\rm Im} \langle \psi_{n,{\bf k}} |p_x| \psi_{n',{\bf k}} \rangle \langle \psi_{n',{\bf k}} |p_y| \psi_{n,{\bf k}} \rangle}{(E_{n',{\bf k}}-E_{n,{\bf k}})^2},\label{eq3}
\end{eqnarray}
where $n$ and $n'$ are the band indices, $\Omega^{z}({\bf k})$ the Berry curvature, $p_x$ ($p_y$) the $x$ ($y$) component of the momentum operator, $\psi_{n,{\bf k}}$ the eigenstate with the eigenenergy $E_{n,{\bf k}}$, and $f(E_{n,{\bf k}},\epsilon)$ the Fermi distribution function for the band $n$ and the wave vector ${\bf k}$ at the energy $\epsilon$ relative to the Fermi level.
In the calculation of $\sigma_{xy}$, the magnetization direction was set to the [001] direction and 729000 k points were used for the Brillouin zone integration ensuring good convergence for $\sigma_{xy}$.
We next calculated the transverse thermoelectric coefficient for a given temperature $T$ using the following expression derived from the Boltzmann transport theory:
\begin{equation}
\alpha_{xy}=-\frac{1}{eT} \int d\epsilon \left( -\frac{\partial f}{\partial \epsilon} \right) (\epsilon-\mu) \sigma_{xy}(\epsilon),\label{eq4}
\end{equation}
where $e>0$ is the elementary charge and $f=1/\{\exp[(\epsilon-\mu)/k_{\rm B}T]+1\}$ is the Fermi distribution function with $\mu$ being the chemical potential. Here, $\mu=0$ corresponds to the Fermi level.

Figure~\ref{Fig1}(c) shows the out-of-plane XRD patterns of the Mn-Ga films.
Besides the diffraction peaks from the MgO substrate, only the sharp (002) and (004) peaks from the Mn-Ga films were observed, indicating an epitaxial growth.
A clear shift of position for both the (002) and (004) peaks toward higher angle with increasing $x$ can be seen.
The in-plane XRD patterns of the Mn-Ga films are shown in Fig.~\ref{Fig1}(d), where the (200) and (400) diffraction peaks can be observed.
From the positions of the diffraction peaks, the lattice constants ($a$ and $c$) of the Mn-Ga films were estimated.
The out-of-plane lattice constant $c$ monotonically decreases with increasing $x$, as shown in Fig.~\ref{Fig1}(e).
Meanwhile, the in-plane lattice constant $a$ shows little dependence on $x$ [Fig.~\ref{Fig1}(f)].
As a results, the tetragonal distortion ratio $c/2a$ of the Mn-Ga films decreases with increasing $x$, with the values roughly between 0.94 and 0.90 [Fig.~\ref{Fig1}(g)].

\begin{figure}
\includegraphics[width=8.6cm]{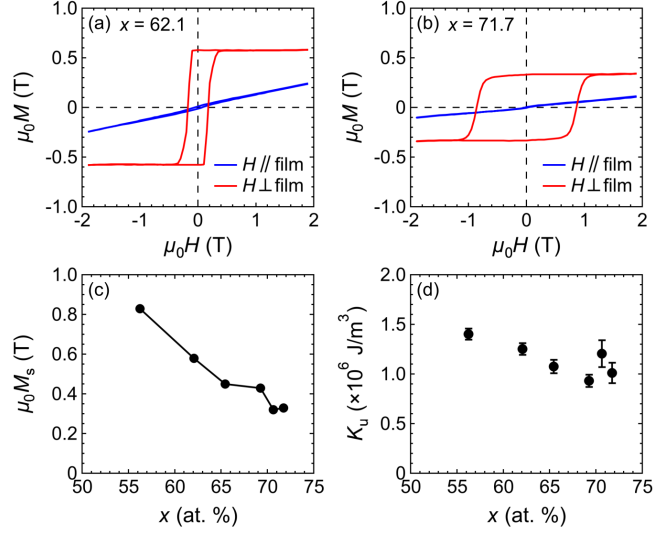}
\caption{\label{Fig2} In-plane and out-of-plane $M$-$H$ curves for the Mn$_{x}$Ga$_{100-x}$ films with (a) $x =$ 62.1 and (b) $x =$ 71.7. (c) Saturation magnetization $M_\text{s}$ and (d) uniaxial magnetic anisotropy $K_\text{u}$ as a function of $x$.}
\end{figure}

The magnetic field ($H$) dependence of magnetization ($M$) for the Mn$_{x}$Ga$_{100-x}$ films with $x =$ 62.1 and 71.7 is shown in Figs.~\ref{Fig2}(a) and \ref{Fig2}(b), respectively.
These $M$-$H$ curves were measured with $H$ applied in-plane ($H \mathrel{/\mkern-5mu/}$ film) or out-of-plane ($H$ $\perp$ film) of the samples.
For all the Mn-Ga films, the magnetization easy axis is out-of-plane.
From the out-of-plane $M$-$H$ curves, $M_\text{s}$ was estimated and plotted as a function of $x$ [Fig.~\ref{Fig2}(c)], which exhibited a clear decrease of $\mu_{0}M_\text{s}$ with increasing $x$ roughly from 0.8 T to 0.3 T.
$K_\text{u}$ of the Mn-Ga films was determined using the relation $K_\text{u} = \frac{\mu_{0}}{2}M_\text{s}H^\text{eff}_{\text{k}} + \frac{\mu_{0}}{2}{M_\text{s}}^2$, where $H^\text{eff}_{\text{k}}$ is the effective magnetic anisotropy field and was estimated using the $M$-$H$ curves measured with $H$ parallel to the magnetization hard axis, i.e., the in-plane $M$-$H$ curves.
All the Mn-Ga films show large values of $K_\text{u}$ roughly in the range between $1.0 \times 10^{6}$ to $1.5 \times 10^{6}$ J/m$^{3}$.
The overall structure and magnetic properties of the Mn-Ga films shown in Figs.~\ref{Fig1} and \ref{Fig2} are consistent with the previously reported results.\cite{MnGa5}

\begin{figure}
\includegraphics[width=8.6cm]{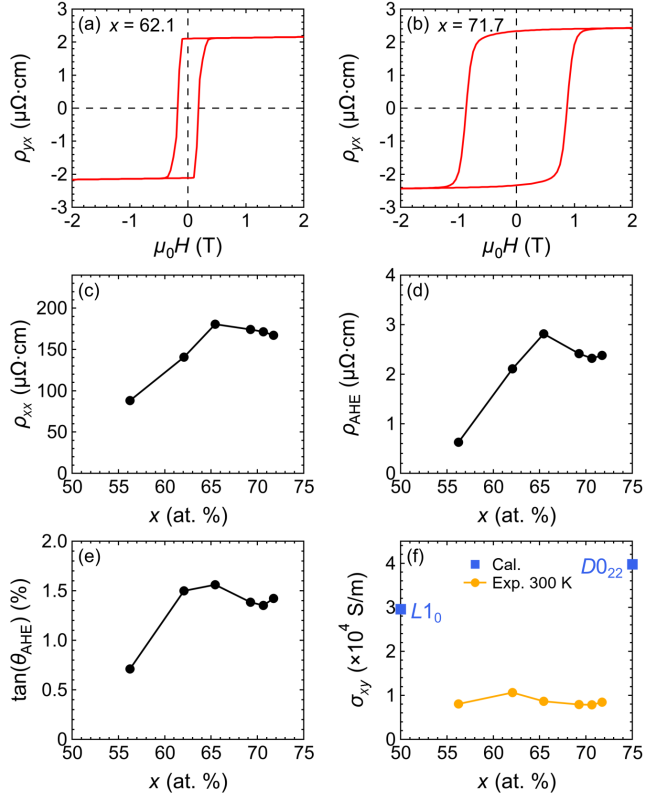}
\caption{\label{Fig3} (a) $H$ dependence of the transverse resistivity $\rho_{yx}$ for the Mn$_{x}$Ga$_{100-x}$ films with $x =$ 62.1 and (b) $x =$ 71.7. $H$ was applied out-of-plane. (c) Longitudinal resistivity $\rho_{xx}$, (d) anomalous Hall resistivity $\rho_\text{AHE}$, (e) anomalous Hall angle $\text{tan}(\theta_\text{AHE})$, and (f) anomalous Hall conductivity $\sigma_{xy}$ as a function of $x$. $\sigma_{xy}$ at the Fermi level obtained from first-principles calculations [Fig.~\ref{Fig5}(a) and \ref{Fig5}(b)] is also shown in (f) for comparison.}
\end{figure}

To separate the $S_\text{I}$ and $S_\text{II}$ terms of the ANE, we measured $\rho_{xx}$ and $\rho_\text{AHE}$ of the patterned Mn-Ga films.
Figures~\ref{Fig3}(a) and \ref{Fig3}(b) show the transverse resistivity ($\rho_{yx}$) as a function of $H$ applied out-of-plane for the Mn$_{x}$Ga$_{100-x}$ films with $x =$ 62.1 and 71.7, respectively.
$\rho_{xx}$ was obtained using 4-terminal method under zero $H$, while $\rho_\text{AHE}$ was obtained by linearly extrapolating the data points of $\rho_{yx}$ at high $H$, where $M$ of the Mn-Ga films was saturated, to zero $H$.
$\rho_{xx}$ and $\rho_\text{AHE}$ as a function of $x$ are shown in Figs.~\ref{Fig3}(c) and \ref{Fig3}(d), respectively.
Figure~\ref{Fig3}(e) shows the anomalous Hall angle ($\text{tan}(\theta_\text{AHE})=\rho_\text{AHE}/\rho_{xx}$) as a function of $x$.
The results of the AHE measurement are in agreement with previous reports.\cite{MnGaAHE1,MnGaAHE2}
The anomalous Hall conductivity ($\sigma_{xy}$) of the Mn-Ga films was determined using the relation $\sigma_{xy}=\rho_\text{AHE}/({\rho_{xx}}^{2}+{\rho_\text{AHE}}^{2})$.
As shown in Fig.~\ref{Fig3}(f), $\sigma_{xy}$ exhibits no significant dependence on $x$.

\begin{figure}
\includegraphics[width=8.6cm]{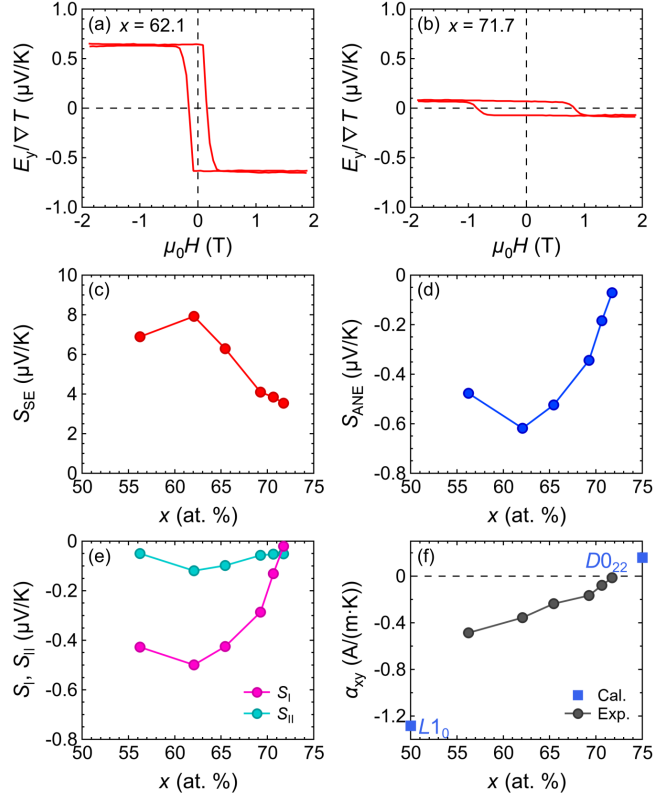}
\caption{\label{Fig4} (a) $H$ dependence of the transverse electric field $E_{y}$ divided by temperature gradient $\nabla{T}$ for the Mn$_{x}$Ga$_{100-x}$ films with $x =$ 62.1 and (b) $x =$ 71.7. $H$ was applied out-of-plane. (c) Seebeck coefficient $S_\text{SE}$, (d) anomalous Nernst coefficient $S_\text{ANE}$, (e) $S_\text{I}$ and $S_\text{II}$ terms of the ANE, and (f) transverse thermoelectric coefficient $\alpha_{xy}$ as a function of $x$. $\alpha_{xy}$ at the Fermi level obtained from first-principles calculations [Fig.~\ref{Fig5}(c) and \ref{Fig5}(d)] is also shown in (f) for comparison.}
\end{figure}

Figures~\ref{Fig4}(a) and \ref{Fig4}(b) show the $H$ dependence of the transverse electric field ($E_{y}$) divided by $\nabla{T}$ for the Mn$_{x}$Ga$_{100-x}$ films with $x =$ 62.1 and 71.7, respectively.
Here, $H$ was applied out-of-plane.
Since $E_{y}$ is due to the ANE of the Mn-Ga films, the curves in Figs.~\ref{Fig4}(a) and \ref{Fig4}(b) exhibit shapes similar to that of the corresponding out-of-lane $M$-$H$ curves in Figs.~\ref{Fig2}(a) and \ref{Fig2}(b), however, reversed along the horizontal axis, due to the negative $S_\text{ANE}$.
The obtained $S_\text{SE}$ and $S_\text{ANE}$ of the Mn-Ga films are summarized in Fig.~\ref{Fig4}(c) and \ref{Fig4}(d), respectively.
For all the Mn-Ga films, $S_\text{SE}$ is positive while $S_\text{ANE}$ is negative.
The value of $S_\text{ANE}$, though, exhibits a dramatic change with different compositions; the largest $|S_\text{ANE}| =$ 0.62 $\mathrm{\mu}$V/K was obtained with $x =$ 62.1, which is close to one order of magnitude larger than the smallest $|S_\text{ANE}| =$ 0.07 $\mathrm{\mu}$V/K with $x =$ 71.7.
This result shows that $|S_\text{ANE}|$ in the Mn-Ga films could be larger than previously considered;\cite{SANEMnGa2} and it also suggests that even in the $L1_{0}$ or $D0_{22}$ ordering phase, $S_\text{ANE}$ could differ a lot depending on the exact composition.
To gain insight into the negative sign and the large difference in $S_\text{ANE}$, we separated the $S_\text{I}$ and $S_\text{II}$ terms of the ANE using Eq.~(\ref{eq1}), and the results are shown in Fig.~\ref{Fig4}(e).
The negative $S_\text{ANE}$ is originated from both the $S_\text{I}$ and $S_\text{II}$ terms being negative.
However, $|S_\text{I}|$ is much larger than $|S_\text{II}|$.
In addition, $S_\text{II}$ shows no significant dependence on $x$, while the dramatic change in $S_\text{ANE}$ is dominated by $S_\text{I}$.
Figure~\ref{Fig4}(f) shows $\alpha_{xy} = S_\text{I} / \rho_{xx}$ as a function of $x$.
Interestingly, $\alpha_{xy}$ is negative and increases monotonically towards zero with increasing $x$.

\begin{figure}
\includegraphics[width=8.6cm]{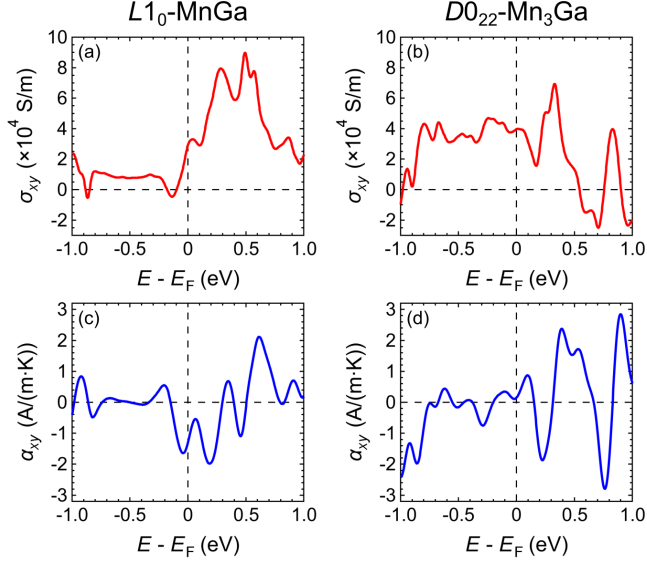}
\caption{\label{Fig5} (a) Anomalous Hall conductivity $\sigma_{xy}$ around the Fermi level ($E_\text{F}$) for $L1_{0}$-MnGa and (b) $D0_{22}$-Mn$_{3}$Ga obtained using the first-principles calculations. (c) Transverse thermoelectric coefficient $\alpha_{xy}$ around $E_\text{F}$ for $L1_{0}$-MnGa and (d) $D0_{22}$-Mn$_{3}$Ga.}
\end{figure}

To understand this behavior, we calculated the theoretical $\alpha_{xy}$ of $L1_{0}$-MnGa and $D0_{22}$-Mn$_{3}$Ga using first-principles calculations.
Figures~\ref{Fig5}(a) and \ref{Fig5}(b) show the calculated $\sigma_{xy}$ around the Fermi level ($E_\text{F}$) for $L1_{0}$-MnGa and $D0_{22}$-Mn$_{3}$Ga, respectively.
The values of $\sigma_{xy}$ at $E_\text{F}$ are also plotted in Fig.~\ref{Fig3}(f) at the corresponding $x$ for comparison.
The signs of $\sigma_{xy}$ from experiment and calculation are in agreement, with the values roughly in the same order, though $\sigma_{xy}$ of $L1_{0}$-MnGa and $D0_{22}$-Mn$_{3}$Ga from calculations is 3 $\sim$ 4 times larger than $\sigma_{xy}$ of the Mn-Ga films obtained in experiment.
It is worth mentioning that 0 K was assumed during the calculation of $\sigma_{xy}$, also the additional electron scattering at the surface and the substrate interface of the sample in thin film form causes the reduction of $\sigma_{xy}$ through the increase of $\rho_{xx}$ in experiment,\cite{axy6,axy10,sigmaxy} which may be attributable to this difference.
Using Eq.~(\ref{eq4}) with $T =$ 300 K, $\alpha_{xy}$ around $E_\text{F}$ was obtained for $L1_{0}$-MnGa and $D0_{22}$-Mn$_{3}$Ga, as shown in Figs.~\ref{Fig5}(c) and \ref{Fig5}(d), respectively.
$\alpha_{xy} =$ $-$1.28 ($+$0.16) A/(m$\cdot$K) at $E_\text{F}$ for $L1_{0}$-MnGa ($D0_{22}$-Mn$_{3}$Ga) is also shown in Fig.~\ref{Fig4}(f).
The calculated $\alpha_{xy}$ matches well with the monotonic increase of $\alpha_{xy}$ with increasing $x$ obtained in experiment.
The results from first-principles calculations also suggest the potential to enhance the ANE in Mn-Ga alloys by Fermi level tuning.
As shown in Fig.~\ref{Fig5}(c) and \ref{Fig5}(d), a dip reaching a larger $|\alpha_{xy}|$ $\sim$ 2 A/(m$\cdot$K) lies at $E - E_\text{F}\sim$ 0.2 eV for both $L1_{0}$-MnGa and $D0_{22}$-Mn$_{3}$Ga.
Using electron doping by adding other elements,\cite{axy10,CMG4} $E_\text{F}$ of the Mn-Ga alloys may be shifted to have larger $|\alpha_{xy}|$, which could lead to enhancement of the negative $S_\text{ANE}$.

In summary, the ANE of Mn-Ga alloys was investigated in experiment by characterizing epitaxial thin films of Mn$_{x}$Ga$_{100-x}$ ($x =$ 56.2, 62.1, 65.4, 69.2, 70.6, and 71.7), and in theory using $\alpha_{xy}$ of $L1_{0}$-MnGa and $D0_{22}$-Mn$_{3}$Ga obtained through first-principles calculations.
The experimentally obtained values of $S_\text{ANE}$ are negative for all samples, and exhibit close to one order of magnitude large difference among different compositions.
The origin of the negative $S_\text{ANE}$ is that both the $S_\text{I}$ and $S_\text{II}$ terms of the ANE are negative, while
the dramatic change in $S_\text{ANE}$ is found to be mainly due to $S_\text{I}$, originated from a monotonically increasing $\alpha_{xy}$ with increasing $x$.
This is consistent with the tendency of the calculated $\alpha_{xy}$ of $L1_{0}$-MnGa and $D0_{22}$-Mn$_{3}$Ga.
By providing a deeper understanding of the negative $S_\text{ANE}$, our results shed light on further enhancement of the ANE in Mn-based ordered alloys, which would promote the development of thermoelectric applications exploiting the ANE in various magnetic materials.

~

We thank K. Uchida, Y. Miura, H. Sukegawa and S. Mitani for valuable discussions.
This work was supported in part by JST PRESTO 'Scientific Innovation for Energy Harvesting Technology' (grant no. JPMJPR17R5) and New Energy and Industrial Technology Development Organization (NEDO) ‘Mitou’ challenge 2050 (grant no. P14004). 

~

The data that support the findings of this study are available from the corresponding author upon reasonable request.

\bibliography{Text_ref}

\end{document}